\documentclass{nature}  

\usepackage{graphicx} 
\usepackage{dcolumn}  
\usepackage{bm}    
\usepackage{amssymb}  
\usepackage{amsmath}
\usepackage{tikz}
\usepackage[justification=justified,format=plain]{caption}

\makeatletter
\let\saved@includegraphics\includegraphics
\AtBeginDocument{\let\includegraphics\saved@includegraphics}
\renewenvironment*{figure}{\@float{figure}}{\end@float}
\makeatother

\mathchardef\mhyphen="2D

\graphicspath{{./FinalFigures/Nature/}}

\DeclareMathOperator{\erf}{erf}

\title{Imaging the nanoscale phase separation in vanadium dioxide thin films at terahertz frequencies}

\author{H. T. Stinson$^{1*}$, A. Sternbach$^{1,2}$, O. Najera$^{3}$, R. Jing$^{1,2}$, A.~S.~Mcleod$^{1,2}$, T.~V.~Slusar$^{4}$, A. Mueller$^{1,5}$, L. Anderegg$^{1,6}$, H. T. Kim$^{4}$, M. Rozenberg$^{1,3}$ \& D. N. Basov$^{1,2}$}

\begin{document}

\maketitle

\begin{affiliations}
\item Department of Physics, University of California, San Diego, La Jolla, California 92093, USA
\item Department of Physics, Columbia University, New York, New York 10027, USA
\item Laboratoire de Physique des Solides, CNRS, Universit\'e Paris-Sud, 91405 Orsay Cedex, France
\item Electronics and Telecommunications Research Institute, Daejeon 34129, South Korea
\item Department of Applied Physics and Materials Science, California Institute of Technology, Pasadena, California 91125, USA
\item Department of Physics, Harvard University, Cambridge, Massachusetts 02138, USA
\end{affiliations}

\begin{abstract}
We use apertureless scattering near-field optical microscopy (SNOM) to investigate the nanoscale optical response of vanadium dioxide (VO\textsubscript{2}) thin films through a temperature-induced insulator-to-metal transition (IMT). We compare images of the transition at both mid-infrared (MIR) and terahertz (THz) frequencies, using a custom-built broadband THz-SNOM compatible with both cryogenic and elevated temperatures. We observe that the character of spatial inhomogeneities in the VO\textsubscript{2} film strongly depends on the probing frequency. In addition, we find that individual insulating (or metallic) domains have a temperature-dependent optical response, in contrast to the assumptions of a classical first-order phase transition. We discuss these results in light of dynamical mean-field theory calculations of the dimer Hubbard model recently applied to VO\textsubscript{2}.
\end{abstract}

Vanadium dioxide (VO\textsubscript{2}) is a canonical example of an insulator-metal transition (IMT) material\cite{morin1959oxides, goodenough1971two} with many exciting potential technological applications\cite{eden1979some, chain1991optical, yang2011oxide}. Although this material has been studied for decades, the precise physics behind its phase transition are still not fully understood. There is an ongoing debate as to whether the transition is a structurally driven Peierls-like one, or whether it is Mott-like and due to electronic correlations\cite{imada1998metal, wentzcovitch1994vo, rice1994comment, cavalleri2004evidence, biermann2005dynamical, koethe2006transfer, kim2006monoclinic, arcangeletti2007evidence}. This complex interplay between Coulomb interaction and structural effects is characteristic of many correlated oxide materials\cite{dagotto2005complexity, basov2011electrodynamics}.

One major difficulty in disentangling this complexity is that conventional experimental methods for examining the phase transition, such as transport or optical spectroscopy, are necessarily insensitive to the state of the material at small length scales\cite{jepsen2006metal, Qazilbash2006}. VO\textsubscript{2} phase-separates into coexisting metallic and insulating domains near the transition temperature $T_{c}$, consistent with the first-order nature of the phase transition\cite{qazilbash2007mott, jones2010nano, liu2013anisotropic}. Here, we use apertureless scattering near-field optical microscopy (SNOM)\cite{Pohl1984a, denk1991near, cvitkovic2007material, amarie2009mid, Novotny2006, atkin2012nano} to investigate the nanoscale optical response through the temperature-driven IMT of VO\textsubscript{2} films grown on sapphire. This technique is able to discriminate between metallic and insulating domains that form near $T_{c}$ in VO\textsubscript{2} thin films\cite{qazilbash2007mott, jones2010nano, liu2013anisotropic}. Our key innovation is to perform this experiment at both mid-infrared (MIR) and terahertz (THz) frequencies, using a custom-built THz-SNOM with a broadband spectrum from 0.1-2 THz (see Methods). THz light interrogates the electronic response at energies on the order of meV, which involve excitations very close to the Fermi level. The ability to investigate the THz response of VO\textsubscript{2} with sub-micron spatial resolution allows us to probe the local low-frequency conductivity of individual domains through the transition.

The principle behind our home-built THz-SNOM is similar in operation to others reported in the literature\cite{von2008spectroscopic, moon2012quantitative, bohmler2016thz}. Our THz-SNOM sample stage is coupled to a variable temperature liquid helium flow cryostat and situated in ultra-high vacuum (Fig.~\ref{fig:schem}a). This allows for measurements from 30-400K with outstanding environmental stability which was essential for the measurements we describe below. The spatial resolution of this THz measurement is approximately 130 nm, limited only by the radius of the custom atomic force microscope (AFM) tips used in our apparatus (see Methods).

\section{Imaging the IMT in VO\textsubscript{2} at THz and MIR frequencies}

In Fig.~\ref{fig:imag}, we show the key experimental data of this work, temperature-dependent images at THz and MIR frequencies of the near-field response of a 100 nm thick VO\textsubscript{2} film grown on sapphire (see Methods). The top row of images are taken at THz frequencies with our custom instrument as described above, and the images in the bottom row are taken in the MIR with a commercial SNOM (Neaspec GmhB) using a 10 $\mu$m CO\textsubscript{2} laser source. We detect the amplitude of the peak of the THz pulse scattered by the tip, demodulated at the second harmonic (S2) due to limited signal to noise. In the the MIR images we report S3, the tip-scattered light demodulated at the third harmonic of the tapping frequency.  Contrast in near-field signal S, which is a measure of local reflectivity, has been shown to reliably discriminate between metallic and insulating regions of VO\textsubscript{2} at MIR frequencies\cite{qazilbash2007mott, jones2010nano, liu2013anisotropic}, and of other spatially inhomogeneous samples at THz frequencies\cite{von2008spectroscopic, moon2012quantitative}. Here we choose a linear color scale with red corresponding to high S metallic regions, and blue to low S insulating regions of the sample. In both the THz and MIR images, the signal S is shown normalized to that over a region of gold in the same field of view. 

The MIR images reveal that upon heating, the VO\textsubscript{2} sample phase-separates into metallic domains within the insulating background. The metallic regions extend through the sample as temperature is increased. This is similar to previous MIR near-field measurements of VO\textsubscript{2}\cite{qazilbash2007mott}. In stark contrast, the THz images appear to evolve homogeneously and continuously from insulating to metallic signal levels through the same temperature region. A histogram representation of the pixel intensity in each image, as shown in Fig.~\ref{fig:meashist}, elucidates this distinction. These histograms exclude the pixels in the gold region of each image. We show histograms of the THz images in Fig.~\ref{fig:meashist}a, and histograms of the MIR images in Fig.~\ref{fig:meashist}b. At temperatures in the middle of the area-averaged transition, the MIR histograms are bimodal. There is an abrupt change in MIR near-field signal between metallic and insulating domains, represented by the separation between the two peaks in the histograms. The pixels in the THz images, on the other hand, are distributed according to a single Gaussian at all temperatures; there is no clear separation in THz near-field signal between insulating and metallic domains. We can track the average THz near-field signal as a function of temperature by fitting the histograms in Fig.~\ref{fig:meashist}a to a single Gaussian distribution and extracting its mean (see Supplementary Note 1 and Supplementary Fig. 1 for details). We plot the mean of each THz histogram as a function of temperature in Fig.~\ref{fig:meashist}c as circles, connected by a dashed line as a guide to the eye. The THz near-field signal appears to evolve continuously with temperature. 

There are two important differences between the THz and MIR SNOM measurements. The first is the spatial resolution; the THz-SNOM employs custom AFM probes which have a larger radius than the MIR probes (see Methods). Although the spatial resolution in our THz-SNOM is coarser than that of the MIR-SNOM, autocorrelation analysis of the THz and MIR images shows that the THz-SNOM resolution is sufficient to resolve the inhomogeneity apparent in the MIR images (Supplementary Note 2 and Supplementary Fig. 2). The second and most relevant difference is the variation in near-field signal level between insulating and metallic end states, which is a function of the probing frequency. As we describe below, the THz images are best explained by assuming that the THz near-field signal close to $T_{c}$ is very similar in the insulating and metallic state. If the relative signal level between these two phases falls below the THz-SNOM noise floor, then the THz images will not be able to resolve the phase boundary. However, an insulating state whose THz near-field signal is close to that of the metallic state at $T_{c}$ implies that the insulating state conductivity is increasing continuously as the temperature approaches $T_{c}$.

A closer examination of the MIR near-field images reveals a similar behavior. We fit the MIR histograms to the sum of two skewed Gaussians, whose means are the insulating and metallic signal levels at each temperature (Supplementary Note 1 and Supplementary Fig. 1). We plot these extracted mean signals in Fig.~\ref{fig:meashist}c as diamonds, of which there are two for each temperature. The larger diamond at each temperature is the center signal of the taller Gaussian, corresponding to the signal level of the majority of pixels (insulating or metallic) in the image. We connect the majority signal levels with a dotted line, which has an abrupt jump in magnitude at the transition temperature. Although this abrupt jump in the MIR near-field signal between insulating and metallic domains is indicative of a first-order transition, we also observe that the MIR near-field signal within the insulating or metallic phase is changing continuously with temperature below and above $T_{c}$. The gradual continuous change in MIR near-field signal of the insulating or metallic state is similar to what we observe in the entire film in the THz, and is contrary to the assumptions of a conventional first-order phase transition.

\section{Nano-THz and nano-infrared contrast in the vicinity of the first order insulator-to-metal transition}

A first-order phase transition implies an order parameter that changes discontinuously with temperature. In the case of an IMT, that order parameter describes the metallicity of the material, and is closely related to the density of states (DOS) at the Fermi level\cite{kotliar2000landau}. In the insulating state the DOS is vanishingly small, and in the metallic state the DOS is finite. As the temperature crosses $T_{c}$, the DOS "jumps" from insulating to metallic. Although the MIR near-field images reveal a jump in near-field signal from insulating to metallic, we also observe a that the near-field signals in the insulating and metallic state are both temperature dependent. In addition, and rather surprisingly, we observe a homogeneous and continuous evolution of the near-field signal at THz frequencies, which probe the electronic response at energies very close to the Fermi level. A traditional first-order IMT does not account for these behaviors. Instead, the continuous change in near-field signal evokes a gradual filling of the DOS in the band gap with increasing temperature. This curious discrepancy calls for resolution.

VO\textsubscript{2} exhibits hysteretic resistance and a divergent molar heat capacity at constant pressure\cite{morin1959oxides, berglund1969electronic}, both indicative of a first-order transition. Thus it seems unlikely that the IMT is truly continuous as suggested by the THz near-field images. Another possibility is that a long-range interaction, such as strain, disorder, or even electronic correlations, leads to a micro-emulsion phase whose characteristic domain size is smaller than the spatial resolution of our near-field measurements\cite{jamei2005universal, spivak2006transport}. This latter line of reasoning disagrees with our own observation of clear domain formation of 100-200nm in the MIR images (Supplementary Note 2 and Supplementary Fig. 2). 

Of course, the ability to resolve separate domains depends not only on spatial resolution, but also on the relative signal level of the two states compared to instrumental signal-to-noise. As mentioned above, we interpret the apparent homogeneity of the THz images as most likely due to a reduced THz contrast between the insulating and metallic state that falls beneath our instrumental noise floor. SNOM is essentially a measure of local reflectivity. In order for the THz reflectivity in the insulating state to be comparable to that of the metallic state, there must be a finite DC conductivity in the insulating state. A small but finite conductivity corresponds to a reflectivity which geometrically approaches 1 as the frequency decreases (Supplementary Note 3). Thus, at THz frequencies even a very small optical conductivity in an insulating state translates into a near-field signal of similar magnitude to that of a metallic state. A finite DC conductivity in the insulating state implies that there is a finite DOS at the Fermi level at temperatures below $T_{c}$. This picture is not inconsistent with far-field THz studies of VO\textsubscript{2}\cite{jepsen2006metal, Qazilbash2009a, cocker2010terahertz}, which observe a gradual increase in THz conductivity below the transition temperature. In previous studies this gradual transition is attributed to the area-averaged nature of the far-field measurement, but many have difficulty in fitting their THz data to effective medium theories. Moreover, this qualitative scenario of a gradual filling of the gap can still be consistent with a first-order transition, according to model calculations of a strongly correlated system relevant for VO\textsubscript{2}.

\section{DMFT solutions of the dimer Hubbard model applied to VO\textsubscript{2} transition}

An explanation for in-gap states prior to the metallic transition has an explicit realization in dynamical mean-field theory (DMFT) solutions of the dimer Hubbard model (DHM) (see Methods). The DHM incorporates structural effects into the Mott-Hubbard Hamiltonian by introducing an additional hopping amplitude $t_{\bot}$, which accounts for the favorability of intra-dimer interaction relevant to the monoclinic structure of VO\textsubscript{2}\cite{goodenough1971two}. The DHM has been recently recognized to capture non-trivial aspects of the IMT in VO\textsubscript{2}, such as the existence of a first-order insulator to metal transition with increasing temperature.\cite{najera2017resolving} Moreover, we show below that for certain values of the ratio $t_{\bot}/U$, the DHM predicts both an abrupt, first-order jump in the density of states at $T_{c}$ and a gradual, almost continuous filling of the gap at temperatures above and below $T_{c}$. These features lead to a qualitative account of our experimental observations.

In Fig.~\ref{fig:spec} we show the frequency- and temperature-dependent DOS, optical conductivity, and simulated SNOM signal for two different values of the ratio $t_{\bot}/U$. Both $t_{\perp}$ and $U$ are given in units of the bandwidth $W=4t$, where $t$ is the interdimer lattice hopping amplitude. For VO\textsubscript{2}, $W\approx 2$eV. The DOS is calculated from DMFT solutions of the DHM, which in turn is used to calculate the optical conductivity. Finally, we use the resulting optical constants to simulate the near-field signal using a lightning-rod model of the tip-sample interaction (see Methods and  Ref.~\citen{mcleod2014model} for details). Panels (a)-(c) correspond to a case with small $t_{\bot}/U$, and panels (d)-(f) are for large $t_{\bot}/U$. Both cases are for values of $t_{\bot}/U$ within the coexistence regime of the DHM phase diagram, meaning that both parameter sets support a first-order temperature-driven IMT with $T_c  \approx 0.02W$ consistent with experiment\cite{najera2017resolving}. However, the temperature dependence of all three plotted quantities is strongly affected by the value of $t_{\bot}/U$.

First we consider the small $t_{\bot}/U$ case, which corresponds to a material with strong electronic correlations compared to the intra-dimer hopping amplitude. In Fig.~\ref{fig:spec}a, the DOS exhibits an abrupt shift from insulating to metallic at a specific temperature $T_{c}$, but does not display strong temperature dependence at temperatures below or above $T_{c}$. This behavior is echoed in the optical conductivity (Fig.~\ref{fig:spec}b); the insulating state has a spectrally flat, insulating conductivity for all temperatures below $T_{c}$, which abruptly jumps to a large Drude-like metallic conductivity at all temperatures above $T_{c}$. In Fig.~\ref{fig:spec}c we show the simulated near-field signal on VO\textsubscript{2} as a function of frequency for different temperatures, which repeats the same general trends as the DOS and the optical conductivity. It is low for $T<T_{c}$ and high for $T>T_{c}$, with an abrupt jump in signal level between the two states at all frequencies. There is very little temperature dependence otherwise.

The temperature dependence is markedly different for the case of large $t_{\bot}/U$, which corresponds to a material with strong dimerization compared to the electronic correlations. In Fig.~\ref{fig:spec}d, there is still an abrupt jump in the DOS at $T_{c}$, but the DOS below and above $T_{c}$ is more temperature dependent. We see that the gap is continuously filling at temperatures below $T_{c}$. This is due to the melting of the intra-dimer singlet at low temperature, whose spectral weight is then spread incoherently over the gap as the temperature increases\cite{najera2017inprep}. The transfer of spectral weight over energy scales that are much higher than $T_{c}$ is a hallmark of strong correlations. In the present context it is related to the competition between the intra-dimer screening of the magnetic moments in the insulator and the lattice Kondo-like screening of each lattice site in the metal state. In other words, this feature can be interpreted as a local RKKY-versus-Kondo screening at the level of a single dimer\cite{degiorgi1999electrodynamic, rozenberg1996transfer, najera2017inprep}. Similarly, in Fig.~\ref{fig:spec}e the optical conductivity at $T<T_{c}$ is still flat and insulating, but is continuously increasing as temperature increases. There is an abrupt jump in conductivity at $T_{c}$ from flat and insulating to Drude-like and metallic, but as temperature continues to rise for $T>T_{c}$ the conductivity continuously increases. For the choice of $t_{\bot}/U$ shown in the bottom row of Fig.~\ref{fig:spec}, the low-frequency conductivity changes by an order of magnitude across the transition. This is smaller than the 3-5 orders of magnitude change in DC resistance observed in transport experiments, and is likely due to the fact that our model does not include a structural transition, but nevertheless is consistent with a first-order IMT.

Notably, the insulating state optical conductivity in the high $t_{\perp}/U$ case has a very different behavior at all frequencies with respect to the low-$t_{\perp}/U$ case. The small but finite low-frequency optical conductivity in the insulating state translates to a 1/$\omega$ behavior in reflectivity, as is expected for a Drude metal whose scattering rate is comparable to the plasma frequency (Supplementary Note 3 and Supplementary Fig. 3). Concurrently, the calculated near-field signal shown in Fig.~\ref{fig:spec}f has a 1/$\omega$-like behavior at low temperatures, with a frequency width that is a function of the DC conductivity $\sigma_0$ and increases with temperature. Thus, the low-frequency near-field signal has a more continuous and smaller relative change with increasing temperature as the conductivity evolves from insulating to metallic. This is reflected in the modeled near-field signal in Fig.~\ref{fig:spec}f, which shows very little temperature variation at low frequencies, but an abrupt jump with temperature at higher frequencies. 

\section{Modeling the VO\textsubscript{2} near-field imaging experiment}

To compare the results of the DHM to our experimental data, we use the calculated frequency-dependent near-field signals in Figs.~\ref{fig:spec}c and \ref{fig:spec}f to simulate a histogram one would obtain in a near-field imaging experiment (Supplementary Note 4). The simulated histograms for the small $t_{\perp}/U$ case are shown in Fig.~\ref{fig:calchist}a. In this case, the temperature dependence of both the THz and MIR near-field signals is similar. The histograms at both frequencies are clearly bimodal. There is a distinct near-field signal level for the insulating and metallic states, with an abrupt jump in signal between the two states. As temperature increases, the near-field signal level in an individual domain does not change, but rather the relative distribution of pixels shifts from majority insulating to majority metallic. The evolution of S(T) for both THz and MIR frequencies is plotted in Fig.~\ref{fig:calchist}b for the small $t_{\perp}/U$ case. At both frequencies, there is an abrupt jump in near-field signal at $T_{c}$, but no temperature dependence above or below $T_{c}$. Thus, the small $t_{\perp}/U$ case is quite different from what we observe experimentally.

The histograms calculated for large $t_{\perp}/U$, which are shown in Fig.~\ref{fig:calchist}c, qualitatively agree with our experimental data. The THz histograms are single Gaussians with a mean near-field signal level that increases continuously with temperature. The MIR histograms are bimodal with an abrupt jump between the two Gaussian centers, yet still display an insulating and metallic near-field signal level that changes continuously with temperature. This is reflected in the temperature-dependent near-field signal at both frequencies, plotted in Fig.~\ref{fig:calchist}d. The THz signal at low temperatures is already very close to the signal at high temperatures, and evolves almost continuously from low to high signal as temperature is increased. The MIR near-field signal, in contrast, has an abrupt jump at $T_{c}$, and is also clearly changing with temperature above and below $T_{c}$. This is similar to what we observe in the experimentally extracted $S(T)$ curves shown in Fig.~\ref{fig:meashist}c.

\section{A modified first-order transition in VO\textsubscript{2}}

Using a novel THz-SNOM with 130 nm spatial resolution, we find that the nature of the domain formation through the phase transition in VO\textsubscript{2} thin films appears homogeneous and continuous at THz frequencies. Moreover, MIR near-field images reveal that the local reflectivity of the insulating or metallic state is changing with temperature below and above $T_{c}$. The DHM appears to provide a framework for understanding how a continuously varying electronic response as revealed by the THz and MIR near-field images can be consistent with a first-order transition. Increased intra-dimer hopping with respect to the Coulomb interaction (large $t_{\perp}/U$) leads to the formation of intra-dimer singlets below a characteristic temperature $T^{*}$\cite{najera2017inprep}. The dissolution of these singlets as $T$ approaches $T_{c}$ results in an incoherent spread of spectral weight across the gap at finite $T<T_{c}$. Gap filling in the insulating state at finite temperature is consistent with previous measurements of both bulk and thin film VO\textsubscript{2}\cite{ladd1969optical, liu2010intrinsic}. This more continuous filling of the DOS for $T<T_{c}$ leads to a small but finite THz conductivity in the insulating state. A small optical conductivity in turn generates a reflectivity which is low but abruptly increases as the frequency approaches 0. Thus, the increased $t_{\perp}/U$ generates a small, finite THz optical conductivity for $T \leq T_{c}$, which translates into an insulating state whose THz reflectivity is larger than the MIR reflectivity. At temperatures very close to the transition, the insulating THz near-field signal is already within the experimental detection limit of the metallic signal. Even though there is an abrupt first-order jump in the optical conductivity at $T_{c}$, the relative change in THz near-field signal across the IMT remains below the experimental detection limit. Therefore, the transition appears homogeneous in the THz images. The transition at $T_{c}$ from insulating to metallic remains abrupt in the MIR near-field signal, but the continuous filling of the gap translates into a continuous shift of the MIR near-field signal in the insulating and metallic state.

As mentioned earlier, the DHM treated within DMFT may also explain the difficulties encountered in fitting low-frequency far-field VO\textsubscript{2} reflection and transmission data to an effective medium theory\cite{Qazilbash2009a, jepsen2006metal, cocker2010terahertz}. According to the DHM, at large values of $t_{\perp}/U$ the low-frequency optical constants of VO\textsubscript{2} thin films are actually changing with temperature in the insulating and metallic states, contrary to the assumptions of effective medium theories. Although there is still an abrupt jump in the DOS at $T_{c}$, the DHM predicts the THz reflectivity in the insulating state just below $T_{c}$ to be very close to the THz reflectivity in the metallic state. 

The key aspect of the DHM is to explicitly incorporate dimerization into the Hamiltonian, thus merging structural effects of the monoclinic state with a correlation-driven IMT. The success of the DHM in explaining our experiments support the conclusion made by others\cite{liu2010intrinsic, biermann2005dynamical, tomczak2008effective} that VO\textsubscript{2} is neither purely a Mott or a Peierls insulator, but a hybrid of the two.

\section{Methods}

\subsection{THz near-field imaging.} We use commercially available LT-GaAs photoconductive antennas (PCAs, Neaspec GmhB)\cite{auston1984picosecond, fattinger1988point} as the THz emitter and detector to form a time-domain spectrometer (TDS)\cite{grischkowsky1990far, tonouchi2007cutting, jepsen2011terahertz}. The PCA emits a broadband pulse with frequencies from 0.1-3 THz. We couple that pulse onto a metallic AFM tip (Rocky Mountain Nanotechnology, LLC) whose length is engineered so as to form an antenna resonance with the peak wavelength of the incoming THz pulse. The resonant tip enhances the field at the apex of the tip and increase tip-sample interaction\cite{huth2013resonant}. THz light is scattered by the tip into the far field, where it is collected and focused onto the detector PCA. Only frequencies up to 2 THz are efficiently scattered by the tip, limiting the bandwidth of the near-field signal to slightly less than that of the PCA. Detection is identical to conventional THz-TDS, except that the detected THz signal is demodulated at higher order harmonics of the tapping frequency of the tip in order to isolate the near-field component of the scattered light\cite{hillenbrand2001pure}. We measure the scattered amplitude of the peak of the THz pulse, corresponding to a frequency-integrated response over the full bandwidth. We show a typical tip-scattered THz pulse in Fig.~\ref{fig:schem}c and the corresponding spectra in Fig.~\ref{fig:schem}d collected over gold and demodulated at the first, second, and third harmonic of the tip tapping frequency. For comparison we also show the far-field (S0) THz pulse and spectrum. 

\subsection{VO\textsubscript{2} film growth}

The VO\textsubscript{2} thin films were fabricated on r-cut Al\textsubscript{2}O\textsubscript{3} substrates by a pulsed laser-deposition method with a 248 nm KrF excimer laser. Prior to the deposition, the chamber was evacuated to a background pressure of $\sim 10^{-6}$ Torr and the substrate was heated up to $600^{\circ}$C. To grow the VO\textsubscript{2} film, a metallic vanadium target was ablated in an oxygen atmosphere at a partial pressure of 30 mTorr. A 30 min deposition process, at the laser pulse energy of 300 mJ and a repetition rate of 10 Hz, yields $\sim100$ nm thick VO\textsubscript{2} films. Au pads were fabricated on the top of the films, using standard photolithography processes and e-beam evaporation. 

\subsection{DMFT calculations of the dimer Hubbard model}
The theoretical calculations were done on a dimer Hubbard model (DHM) solved within cluster-Dynamical Mean Field Theory (C-DMFT). This model consists of dimers on each unit cell with intra-dimer hopping, in addition to the standard inter-dimer hopping and a Hubbard-type local Coulomb repulsion. The model Hamiltonian is given by
\begin{equation}
H=\Biggl[-t\sum_{\langle i, j\rangle \alpha \sigma}c_{i\alpha\sigma}^{\dagger}c_{j\alpha\sigma} +t_{\perp}\sum_{ i \sigma}c_{i1\sigma}^{\dagger}c_{i2\sigma}+\textrm{H.c.}\Biggr]+\sum_{ i \alpha}Un_{i\alpha\uparrow}^{\dagger}n_{i\alpha\downarrow}
\end{equation}
where $\langle i, j\rangle$ denotes nearest-neighbor sites, $\alpha=\lbrace 1,2 \rbrace$ denote the dimer orbitals, $\sigma$ is the spin, $t$ is the inter-dimer (lattice) hopping amplitude, $t_{\perp}$ is the intra-dimer hopping amplitude, and $U$ is the Coulomb repulsion. The DHM was recently shown to capture a thermally driven insulator-to-metal transition for parameters relevant to VO\textsubscript{2}\cite{najera2017resolving}. It also provides a consistent description of the near field optical conductivity data across the IMT in that compound.

The model is defined on a semi-circular non-interacting density of states, which is realized on a Bethe lattice\cite{moeller1999rkky}. The bandwidth of the model is $W=2D=4t$. The qualitative behavior of the model is not strongly affected by the specific type of lattice adopted. The C-DMFT equations are solved within the Iterated Perturbation Theory approximation. At half filling, this approximation is excellent. It was found to be asymptotically exact in many cases, including the weak interacting limit ($U\rightarrow0$) and in the atomic limit ($t\rightarrow0$) for all values of the inter-dimer hopping $t_\perp$. We have extensively benchmarked the approximation against exact but more numerically costly quantum Monte Carlo calculations. Key for our present study, we have implemented a novel finite temperature and real frequency impurity solver, which avoids the technical difficulties of analytic continuation. This allows us to obtain the detailed evolution of the entire density of states with temperature with unprecedented precision.

\subsection{Lightning-rod model calculations of near-field signals}

For modeling of our near-field data we employed the lightning-rod (LR) model of probe-sample near-field interaction\cite{mcleod2014model}. The Fresnel reflection coefficient $r_{p}$ of the sample for light polarized parallel to the tip axis determines the near-field signal and is evaluated for a VO\textsubscript{2} single crystal. By formulating the quasi-electrostatic near-field interaction as a scattering problem in momentum space, the LR model generally provides excellent quantitative agreement with near-field spectroscopy measurements at very low computational cost\cite{mcleod2014model}. Experimental details, such as scattering of light from the probe to the detector and demodulation of the detected signal, are included explicitly in the model. Input parameters include the tip radius $a$ and tapping amplitude $A$, as well as the overall probe geometry, modeled here as a metallic cone 19 $\mu$m in height with half-angle $\approx30^{\circ}$, in qualitative accordance with the geometry of commercial probes used for this study. Best results were obtained with $a=100$ nm and $A=$200 nm, which agree well with nominal experimental values of $a\approx140$ nm and $A\approx250$ nm.

\section{Acknowledgements}

We acknowledge the valuable discussions about optical design and principles of operation for customized terahertz instruments with Dr. Ian Gregory of TeraView Limited, Cambridge.

M.R. acknowledges support by public grants from the French National Research Agency (ANR), project LACUNES No. ANR-13-BS04-0006-01, and the French-US Associated International Laboratory on Nanoelectronics funded by CNRS.

Work at UC-San Diego and Columbia University is supported by ARO-W911nf-17-1-0543. D.N.B. is the the Gordon and Betty Moore Foundation’s EPiQS Initiative investigator, Grant GBMF4533.

\section{Author contributions}

D.B., H.S., A.S., T.S., and H.K. conceived the experiment. H.S. performed the THz near-field imaging experiments. H.S. and A.S. performed the MIR near-field imaging experiments. H.S., R.J., A.S.M., A.M., and L.A. contributed to the design and construction of the THz near-field microscope. T.S. and H.K. grew the VO\textsubscript{2} samples. O.N. and M.R. conducted the DMFT studies on the dimer Hubbard model. All authors prepared the manuscript.

\section{Competing financial interests}

The authors declare no competing financial interests.

\subsection{Code Availability}

The packages used for calculating DMFT solutions are available from M.R. upon reasonable request. The packages used for calculating the lightning-rod model of the near-field signal are available from the corresponding author upon reasonable request.

\subsection{Data Availability}

The data supporting the findings of this work are available from the corresponding author upon reasonable request.

%

\clearpage

\begin{figure}
\centering
\includegraphics{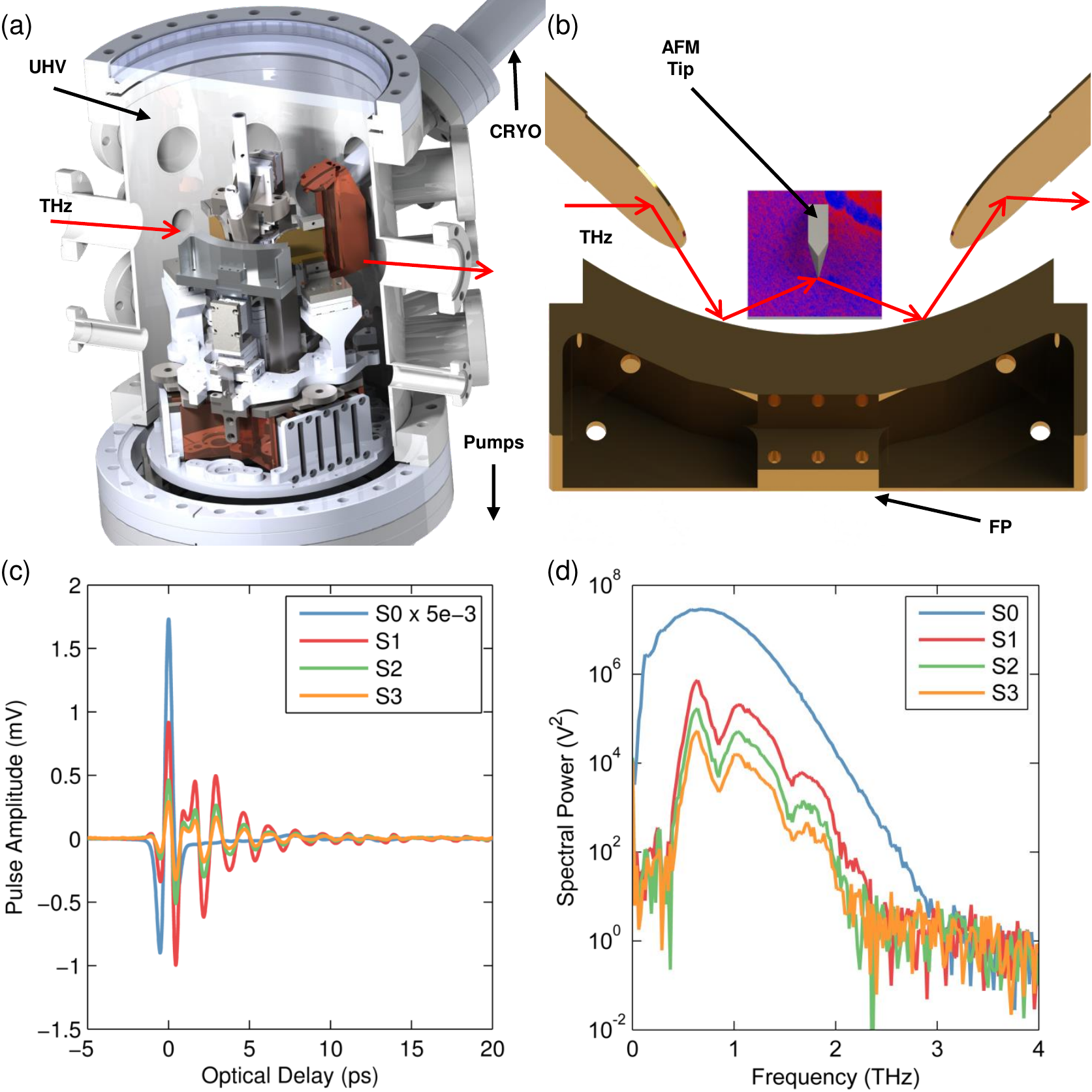}
\end{figure}
\clearpage
\captionof{figure}{(a) Schematic of the THz-SNOM. The THz pulse (red arrow) is sent through a UHV vacuum chamber housing the custom SNOM. The sample is coupled to a heater and liquid-He flow cryostat (CRYO), allowing for operation at temperatures from 40K-400K. (b) Detail view of the SNOM inside the chamber. The same focusing parabola (FP) is used to both focus the THz pulse onto the tip (not shown to scale) and collect the tip-scattered light. (c) Broadband THz pulse (blue, S0) and the near-field THz pulse measured in the system on gold in a dry-air-purged environment. S1, S2, and S3 are the detected THz signal demodulated at the first, second, and third harmonic of the tip tapping frequency. (d) Measured THz spectrum of the far-field pulse (blue) and the near-field spectrum for different harmonics, collected on gold.} 
\label{fig:schem}
\clearpage

\begin{figure}
\includegraphics{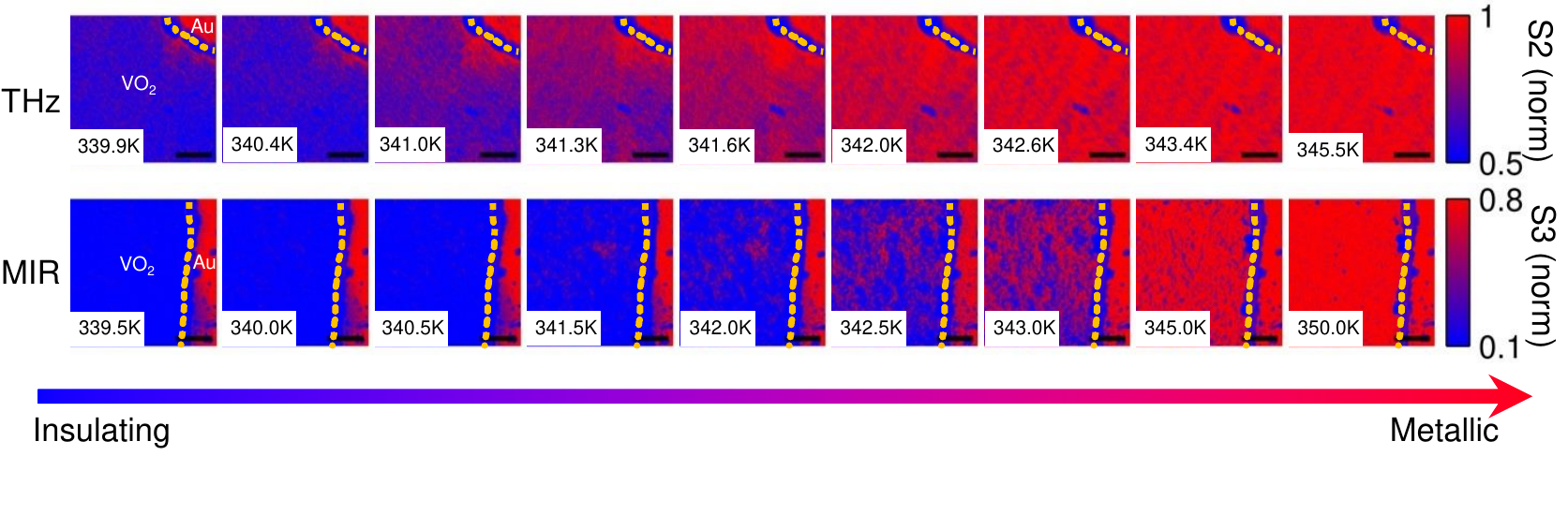}
\end{figure}
\captionof{figure}{(Color online.) SNOM images of the VO\textsubscript{2} IMT taken at THz (top row) and MIR (bottom row) frequencies. The temperature of each image is noted in the bottom left corner. In all images, the signal at every temperature is normalized to the average signal obtained on gold (bright red region in the upper right or right of the image, for THz and MIR respectively). The dashed yellow line denotes the boundary between VO\textsubscript{2} and gold regions. The THz and MIR data are S2 and S3, which is the detected signal demodulated at the second and third harmonic of the tip tapping frequency, respectively. Low near-field signal (blue) is measured in insulating regions, while high near-field signal (red) corresponds to a metallic state. The scales are different for the THz and MIR images to highlight the transition from insulator to metal in both cases. Scale bar, 2 $\mu$m.}
\label{fig:imag}

\clearpage

\begin{figure}
\includegraphics{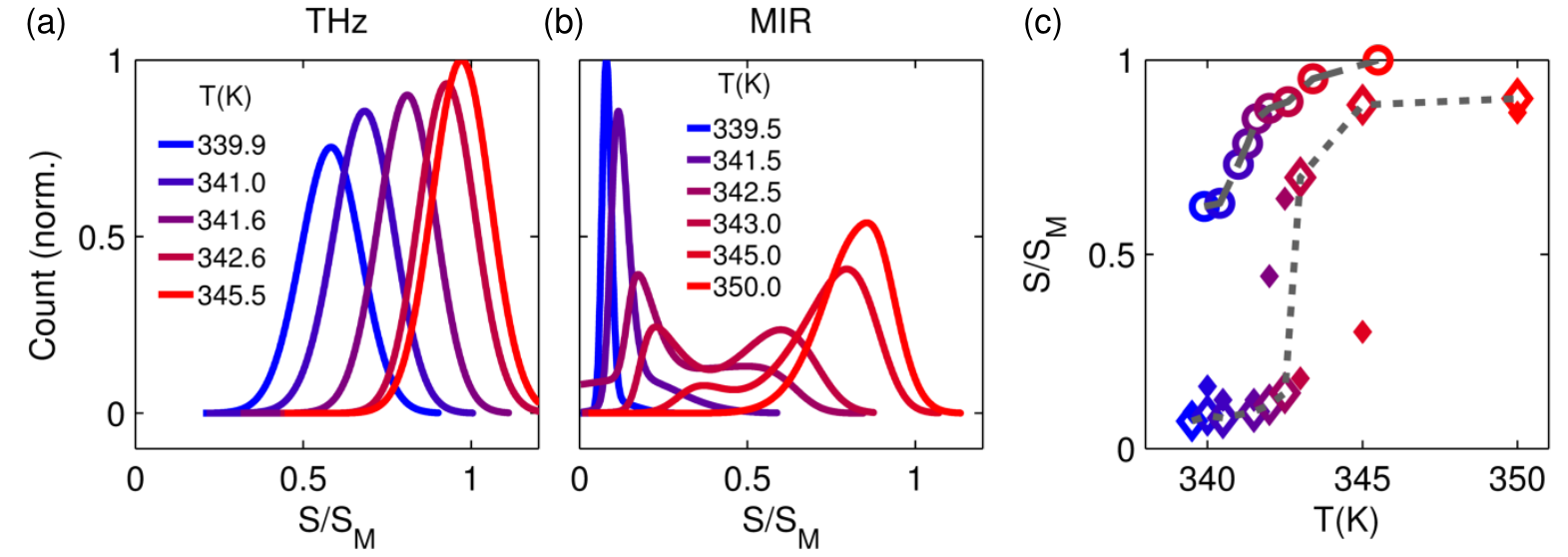}
\end{figure}
\captionof{figure}{(Color online.) Pixel intensity histograms of selected THz (a) and MIR (b) images shown in Fig.~\ref{fig:imag}. (c)~Peak signal level as a function of temperature in the THz (circles) and MIR (diamonds) extracted from single or bi-modal Gaussian fits to the histograms. In the MIR case, there are two peaks at each temperature due to the bimodal nature of the pixel intensity distribution. The maximum peak at each temperature is an open diamond, while the smaller peak is a filled diamond (Supplementary Note 1). The dashed (THz) and dotted (MIR) lines are guides to the eye.}
\label{fig:meashist}

\clearpage

\begin{figure}
\includegraphics{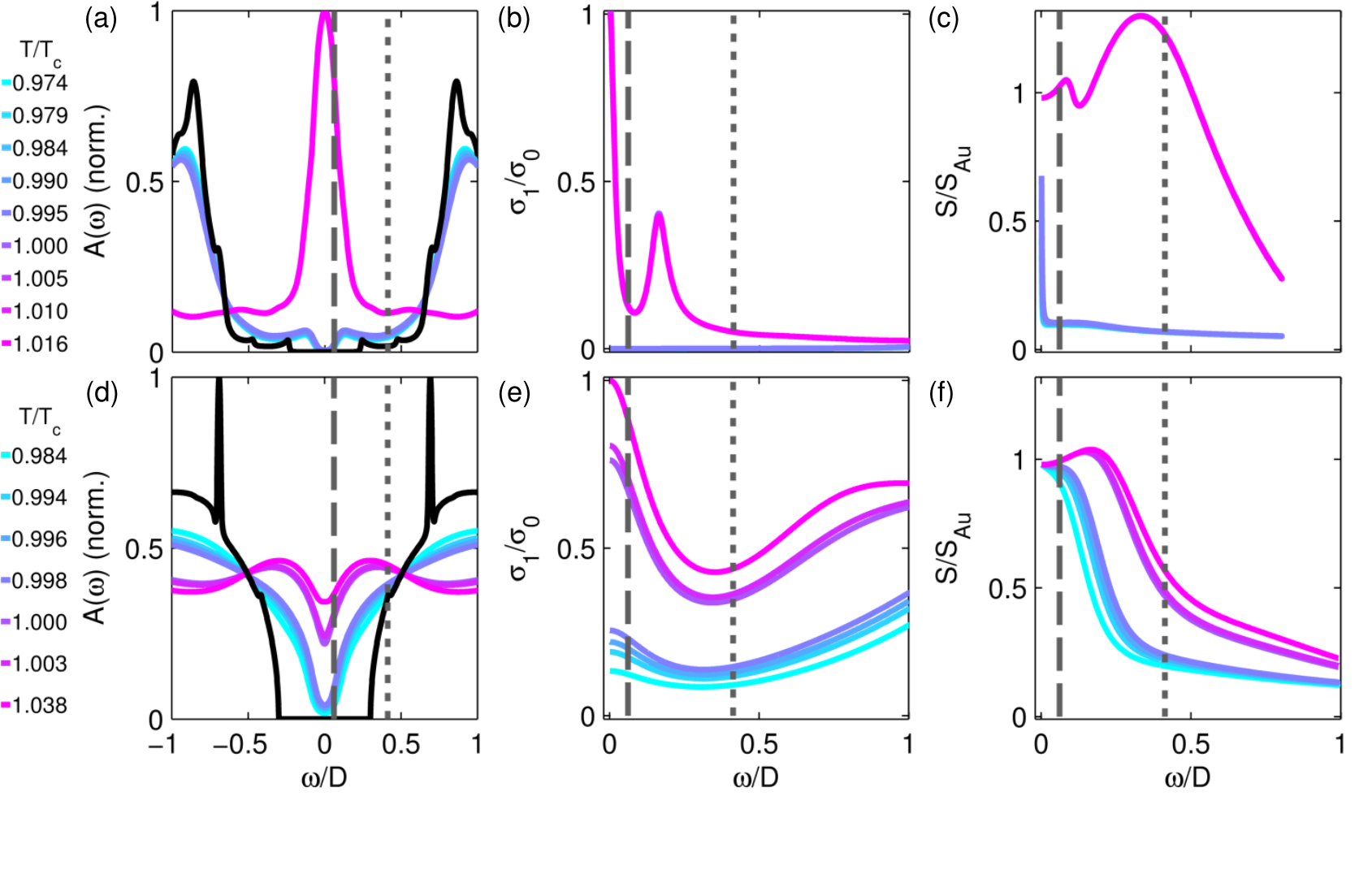}
\end{figure}
\captionof{figure}{(Color online.) Spectra at different temperatures around $T_{c}$, calculated from DMFT for the dimer Hubbard model with parameters $t_{\perp}=0.2, U=3.1$ (a-c) and $t_{\perp}=0.5, U=2.405$ (d-f) as a function of energy normalized to the half-bandwidth D. The temperatures (normalized to $T_{c}$) of each curve for small $t_{\perp}/U$ (a-c) are shown to the left of panel (a). The temperatures used for large $t_{\perp}/U$ (d-f) are shown to the left of panel (d). The left column shows the local density of states (LDOS) at different temperatures as a function of energy for small $t_{\perp}/U$ (a) and large $t_{\perp}/U$ (d). The black line is the LDOS at $T=0$. The sharp peak in the $T=0$ LDOS in panel d is due to the formation of an intra-dimer singlet at very low temperatures (see text for details). The middle column shows the real part of the optical conductivity at different temperatures for small $t_{\perp}/U$ (b) and large $t_{\perp}/U$ (e). All conductivities are normalized to the DC conductivity at the highest temperature shown (i.e., the DC conductivity of the metallic state). The right column shows the calculated near-field signal at different temperatures as a function of frequency for small $t_{\perp}/U$ (c) and large $t_{\perp}/U$ (f). The vertical grey lines in all figures indicate the THz (dashed) and MIR (dotted) frequencies used for calculating histograms in Fig.~\ref{fig:calchist}.}
\label{fig:spec}
\clearpage

\begin{figure}
\includegraphics{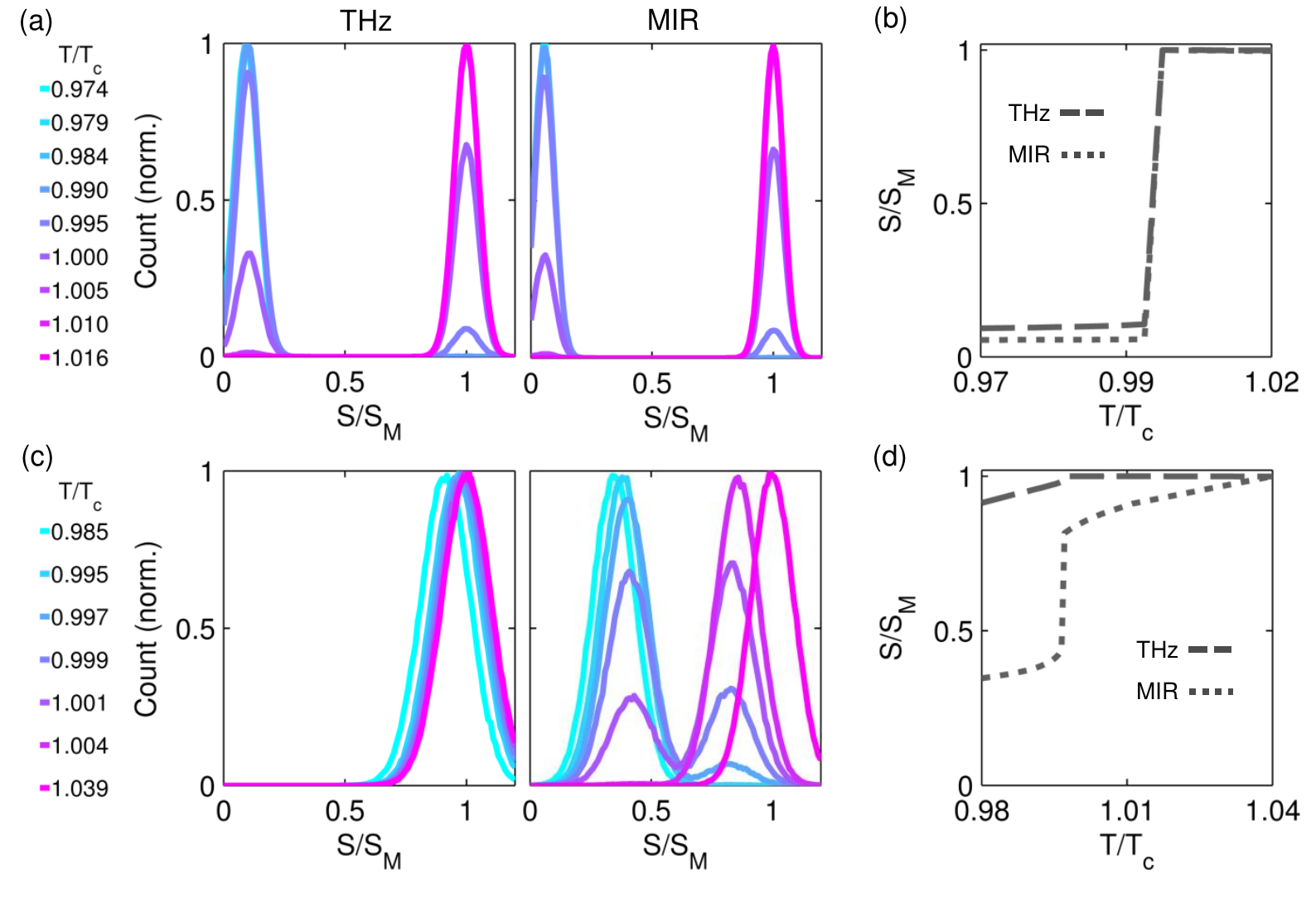}
\end{figure}
\captionof{figure}{(Color online.) Histograms of a phase-separated image using near-field signals at different temperatures calculated from the DHM. See text for details. (a) The simulated histograms for a THz (left) and MIR (right) image in the case of small $t_{\perp}/U$. The signals are calculated at the frequencies represented by the dashed (THz) and dotted (MIR) vertical lines in Fig.~\ref{fig:spec}c. The temperatures shown are normalized to $T_{c}$. (b) Plot of the calculated near-field signal as a function of temperature for small $t_{\perp}/U$ at THz (dashed) and MIR (dotted) frequencies. (c) Same as (a), but for large $t_{\perp}/U$. The signals are calculated at the frequencies represented by the dashed (THz) and dotted (MIR) vertical lines in Fig.~\ref{fig:spec}f. (d) Same as (b), but for large $t_{\perp}/U$. \label{fig:calchist}}

\clearpage
\renewcommand{\thefigure}{S\arabic{figure}}
\setcounter{figure}{0}

\section*{Supplementary Material for "Imaging the phase separation in vanadium dioxide thin films at terahertz frequencies"}

\setcounter{section}{0}

\section{Gaussian fits to near-field image histograms}

First we obtain histograms of the near-field signal distribution in each THz and MIR image. We crop each image to a 210x125 pixel region that does not contain any of the reference gold. We then split the pixel values into 100 separate bins ranging from the minimum to the maximum pixel signal in each image. Raw histograms are plotted as open symbols in Fig.~\ref{fig:histfit} for three representative THz (Fig.~\ref{fig:histfit}a) and MIR (Fig.~\ref{fig:histfit}b) images. As mentioned in the main text, the THz histograms are Gaussian at every temperature, while the MIR histograms are bimodal for intermediate temperatures.

To extract an average signal level at each temperature we fit the THz histograms to the skewed Gaussian function

\begin{equation}
Ae^{-\frac{\left(x-\mu\right)^{2}}{2\sigma^{2}}}\left( 1-\erf{\left[\frac{ \alpha\left( x-\mu\right)}{\sqrt{2}\sigma}\right]}\right)
\label{eq:thzfit}
\end{equation}

where the four free parameters are the amplitude $A$, the standard deviation $\sigma$, the skew parameter $\alpha$, and the mean $\mu$. The fitted functions are plotted as solid lines for three representative temperatures in Fig.~\ref{fig:histfit}a. This last parameter $\mu$ is what is plotted as a function of temperature in Fig. 3c (open circles) of the main text.

Because the MIR images display phase separation, the histograms are bimodal. Therefore we fit the MIR histograms to a function for the sum of two skewed Gaussians:

\begin{equation}
A_{1}e^{-\frac{\left(x-\mu_{1}\right)^{2}}{2\sigma_{1}^{2}}}\left( 1-\erf{\left[\frac{ \alpha_{1}\left( x-\mu_{1}\right)}{\sqrt{2}\sigma_{1}}\right]}\right)+A_{2}e^{-\frac{\left(x-\mu_{2}\right)^{2}}{2\sigma_{2}^{2}}}\left( 1-\erf{\left[\frac{ \alpha_{2}\left( x-\mu_{2}\right)}{\sqrt{2}\sigma_{2}}\right]}\right)\label{eq:mirfit}
\end{equation}

The fitted functions are plotted as solid lines in Fig.~\ref{fig:histfit}b for three representative temperatures. Similarly to the THz data, we plot $\mu_{1}$ and $\mu_{2}$ in Fig. 3c (diamonds) of the main text. The open diamonds have the same subscript as the maximum of $A_{1}$ or $A_{2}$; that is, the open diamonds are the average signal value for the majority of pixels in the image. The filled diamonds are the average signal value of the remaining Gaussian component. 

\section{Autocorrelation analysis of AFM resolution}

The custom AFM tips used for THz-SNOM limit the AFM resolution in THz measurements. Fig.~\ref{fig:snomres}a shows a typical AFM image of the VO\textsubscript{2} film taken with the THz-SNOM. The grains in the film appear larger than in the MIR-SNOM (Fig.~\ref{fig:snomres}b) due to the increased radius of the THz-SNOM tip. We quantify the difference in AFM resolution by performing an autocorrelation analysis of the two AFM images. The full width at half maximum (FWHM) of the autocorrelation curve determines the average feature size of an AFM image. The THz AFM has an autocorrelation FWHM of 134nm (Fig.~\ref{fig:snomres}d, red line), whereas the MIR AFM reveals an average VO\textsubscript{2} grain size of 50nm. 

Fig.~\ref{fig:snomres}c show a MIR near-field image of metallic domains in an insulating background at 342K. These domains are composed of many individual grains. A similar analysis of this image obtains an autocorrelation peak FWHM of 106nm (Fig.~\ref{fig:snomres}d, black line), which we can assign to the average size of metallic domains. Thus, the THz-AFM has sufficient resolution to resolve the metallic domains at some point in the transition.

\section{Low-frequency reflectivity of an incoherent metal}

The Drude model provides a consistent basis for discussing the frequency dependence of material optical response. The dielectric function $\tilde{\epsilon}=\epsilon_{1}+i\epsilon_{2}$ is given by the Drude model as

\begin{align}
\epsilon_{1}=1-\omega_{P}^{2}\tau^{2}\frac{1}{1+\omega^{2}\tau^{2}}\\
\epsilon_{2}=\frac{\omega_{P}^{2}\tau}{\omega}\frac{1}{1+\omega^{2}\tau^{2}}
\end{align}

where $\omega_{P}$ and $\tau$ are the plasma frequency and scattering time, respectively. For simplicity we consider the reflectivity of a material in vacuum at normal incidence, which is

\begin{equation}
R=\left\lvert\frac{\tilde{n}-1}{\tilde{n}+1}\right\rvert
\label{eq:r}
\end{equation}

where $\tilde{n}=n+ik$ is the complex index of refraction of the material. When the magnetic permiability of the material is equal to 1, the index of refraction is related to the optical permitivitty as follows:

\begin{align}
n^{2}=\frac{\sqrt{\epsilon_{1}^{2}+\epsilon_{2}^{2}}+\epsilon_{1}}{2}\\
k^{2}=\frac{\sqrt{\epsilon_{1}^{2}+\epsilon_{2}^{2}}-\epsilon_{1}}{2}.
\end{align}

Consider the case of an incoherent metal where $\omega_{P}\tau\approx1$ in the low frequency limit $\omega\tau\ll1$. This represents a material with a low but finite conductivity, as for the high $t_{\perp}/U$ case in the insulating state (main text). In this limit we obtain the following approximations:

\begin{align}
\epsilon_{2}\approx\frac{\omega_{P}^{2}\tau}{\omega} \label{eq:eps}\\
n\approx q \approx \frac{\sqrt{\epsilon_{2}}}{2}
\label{eq:nq}
\end{align}

Substituting Eqs.~\ref{eq:eps} and \ref{eq:nq} into Eq.~\ref{eq:r}, we obtain

\begin{equation}
R\approx\frac{\dfrac{\omega_{P}}{\omega}-\sqrt{2\dfrac{\omega_{P}}{\omega}}+2}{\dfrac{\omega_{P}}{\omega}+\sqrt{2\dfrac{\omega_{P}}{\omega}}+2}
\label{eq:modref}
\end{equation}

This function is plotted in Fig.~\ref{fig:modref} for $\omega_{P}=1/\tau=5000 \textrm{cm}^{-1}$ along with the full Drude expression for reflectivity. At low frequencies, the reflectivity begins to increase rapidly, approaching 1 for $\omega=0$. 

\section{Generating near-field histograms from dimer Hubbard model calculations}

Phase separation in granular VO\textsubscript{2} films is likely due to local variations in $T_{c}$ caused by long-range interactions such as strain, local stoichiometric variations, or long-range Coulomb interactions\cite{jamei2005universal, mcleod2016nanotextured}. This is equivalent to having a material where $T_{c}$ is homogenous, but there are local variations in the temperature of the film. We use the calculated near-field signal as a function of temperature to model the near-field signal distribution that would be acquired in an imaging experiment. For each temperature $T_{j}$ we generate an $N$-element vector $V$ of temperatures that are normally distributed about $T_{j}$ with a standard deviation of 1K. In Figs. 5a and 5c of the main text, we plot the histogram of $S(V)+\delta S$, where $S(x)$ is the near-field signal calculated at temperature $x$, and $\delta S$ is a random white noise term. The white noise is added at each pixel to simulate the experimental error in detected near-field signal due to electronic and environmental background.
\clearpage

\section{Supplementary References}

\clearpage
\begin{figure}
\centering
\includegraphics{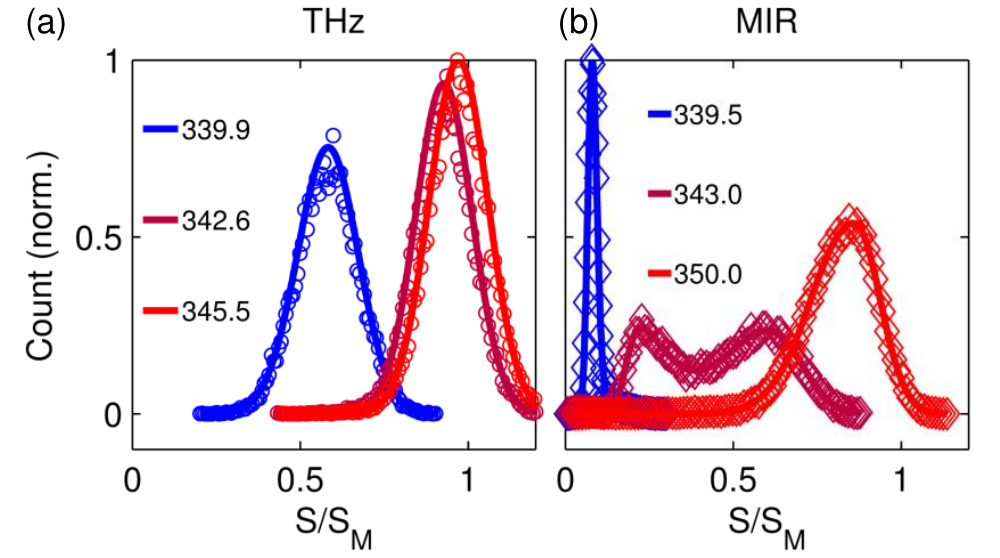}
\end{figure}
\captionof{figure}{Three representative histograms of near-field images taken at THz (a) and MIR (b) frequencies. The open symbols are the raw histogram bin heights, and the solid lines are best fits to the functions outlined in the text.}
\label{fig:histfit}
\clearpage

\clearpage
\begin{figure}
\centering
\includegraphics{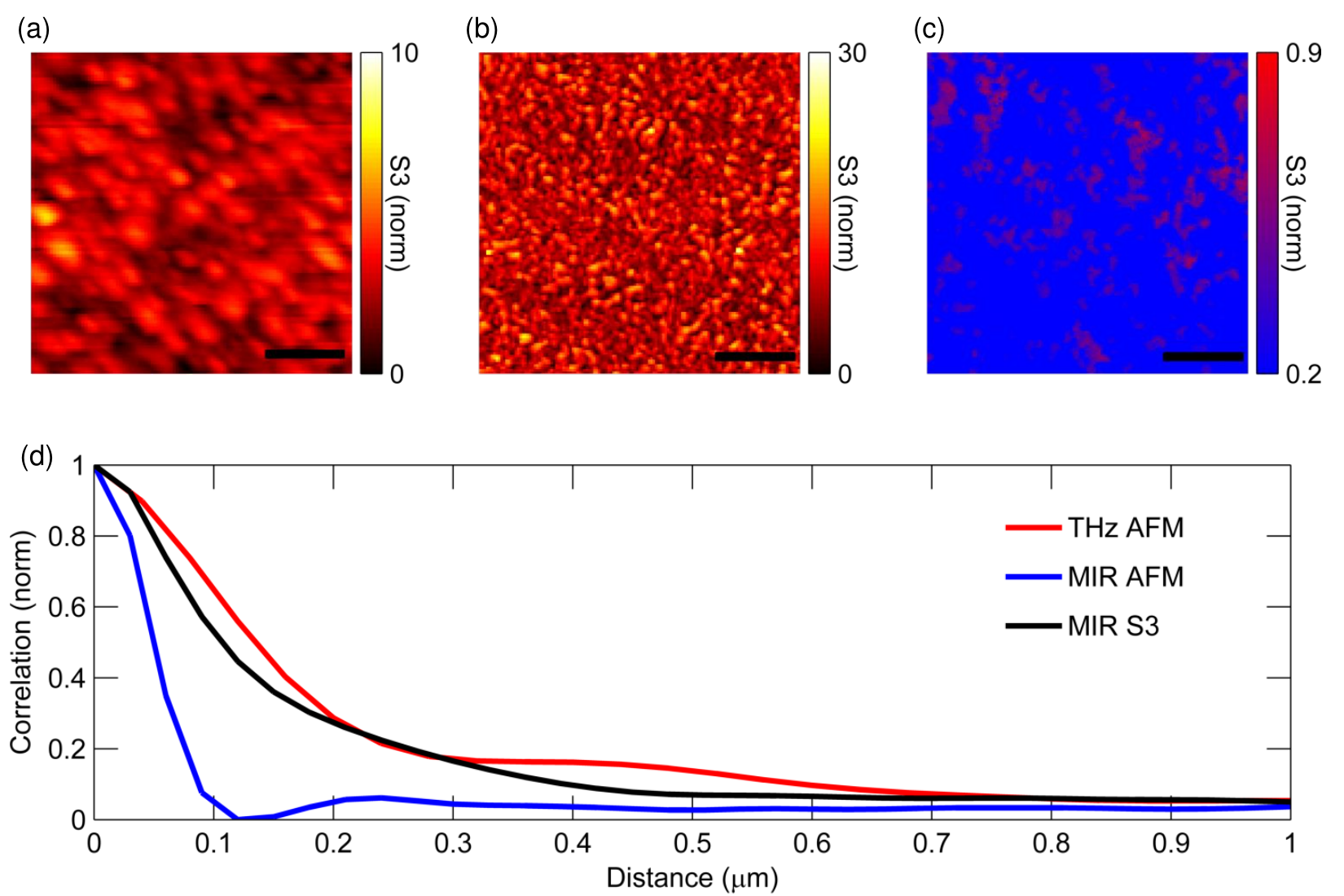}
\end{figure}
\captionof{figure}{Analysis of the THz-SNOM spatial resolution. (a) THz-SNOM AFM image at 342 K. (b) MIR-SNOM AFM image at 342K. (c) MIR-SNOM S3 image of VO\textsubscript{2} metallic domains at 342K. Scale bar in all images is 1 $\mu$m. (d) 2D autocorrelation curves for the three images. The width of the peak at the center of the autocorrelation corresponds to the size of features in the images.}
\label{fig:snomres}

\clearpage

\begin{figure}
\centering
\includegraphics[scale=1]{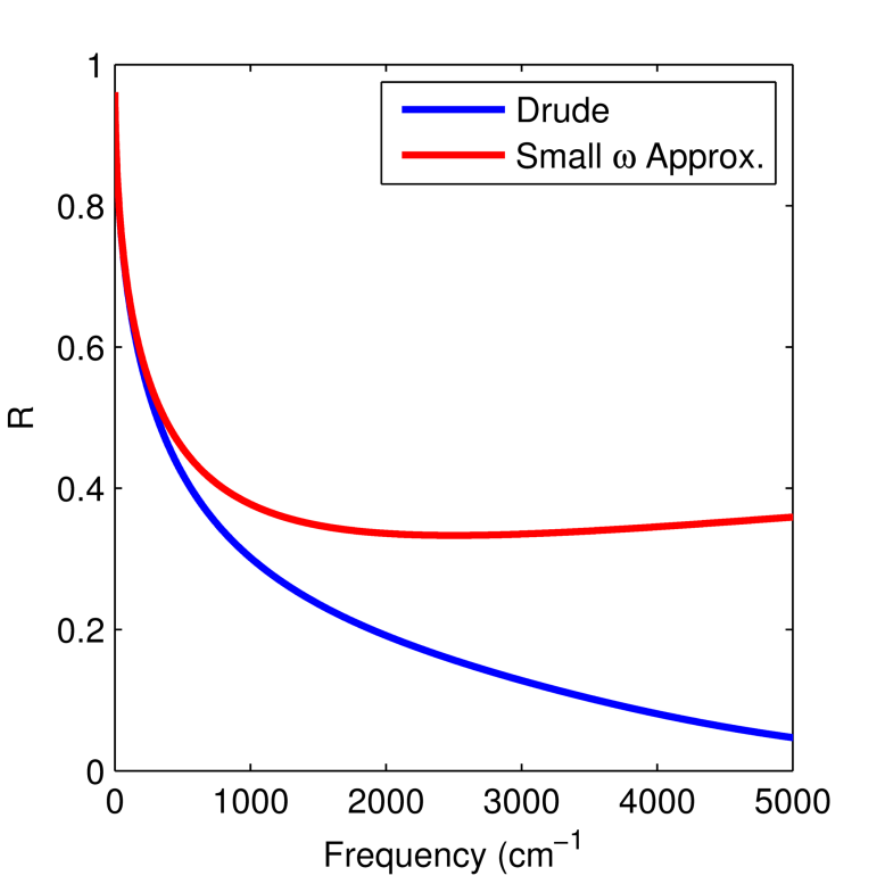}
\clearpage
\captionof{figure}{Theoretical reflectivity of a Drude material with $\omega_{P}=1/\tau=5000\textrm{cm}^{-1}$ as calculated with the full Drude model or with Eq.~\ref{eq:modref}.}
\label{fig:modref}
\end{figure}
\clearpage

\end{document}